\documentclass[a4paper,conference]{IEEEtran}
\IEEEoverridecommandlockouts

\usepackage{cite}
\usepackage{amsmath,amssymb,amsfonts}
\usepackage{algorithmic}
\usepackage{algorithm}
\usepackage{graphicx}
\usepackage{textcomp}
\usepackage{xcolor}
\usepackage{multirow}
\usepackage{booktabs}
\usepackage{pgfplots}
\usepackage{tikz}
\usepackage{subcaption}
\usepackage{tcolorbox}
\pgfplotsset{compat=1.17}

\def\BibTeX{{\rm B\kern-.05em{\sc i\kern-.025em b}\kern-.08em
    T\kern-.1667em\lower.7ex\hbox{E}\kern-.125emX}}

\begin{document}

\title{Hierarchical Multi-Modal Threat Intelligence Fusion Without Aligned Data: A Practical Framework for Real-World Security Operations}

\author{
\IEEEauthorblockN{Sisir Doppalapudi}
\IEEEauthorblockA{\textit{School of Computing and Augmented Intelligence} \\
\textit{Arizona State University}\\
Tempe, AZ, USA \\
sdoppal2@asu.edu}
}

\maketitle

\begin{abstract}
Multi-modal threat detection faces a fundamental challenge that involves security tools operating in isolation, and this creates streams of network, email, and system data with no natural alignment or correlation. We present Hierarchical Multi-Modal Threat Intelligence Fusion (HM-TIF), a framework explicitly designed for this realistic scenario where naturally aligned multi-modal attack data does not exist. Unlike prior work that assumes or creates artificial alignment, we develop principled methods for correlating independent security data streams while maintaining operational validity. Our architecture employs hierarchical cross-attention with dynamic weighting that adapts to data availability and threat context, coupled with a novel temporal correlation protocol that preserves statistical independence. Evaluation on UNSW-NB15, CSE-CIC-IDS2018, and CICBell-DNS2021 datasets demonstrates that HM-TIF achieves 88.7\% accuracy with a critical 32\% reduction in false positive rates—even without true multi-modal training data. The framework maintains robustness when modalities are missing, making it immediately deployable in real security operations where data streams frequently have gaps. Our contributions include: (i) the first multi-modal security framework explicitly designed for non-aligned data, (ii) a temporal correlation protocol that avoids common data leakage pitfalls, (iii) empirical validation that multi-modal fusion provides operational benefits even without perfect alignment, and (iv) practical deployment guidelines for security teams facing heterogeneous, uncoordinated data sources.
\end{abstract}


\section{Introduction}

Security operations centers (SOCs) face a paradoxical challenge: they deploy multiple detection systems—network monitors, email filters, endpoint agents—yet these tools operate in complete isolation. A phishing email that leads to network infiltration appears as separate, uncorrelated events in different systems. This fundamental disconnect between security data sources represents one of the most significant operational challenges in modern cybersecurity \cite{b1}. The core problem is not the lack of sophisticated attacks spanning multiple vectors—these certainly exist. Rather, it is that naturally aligned multi-modal security data essentially does not exist in practice. Available datasets are collected independently: network traffic captures like UNSW-NB15 contain no email data, phishing corpora like IWSPA-AP include no network traces, and system logs remain proprietary and disconnected. This reflects the reality of security operations where different tools, vendors, and teams manage disparate data streams without coordination.

Prior multi-modal security research has largely sidestepped this challenge by creating artificial alignments or assuming correlations that don't exist in practice \cite{b8,b9,b10}. While these approaches demonstrate theoretical possibilities, they fail to address the practical reality faced by security teams: how to effectively combine multiple security data streams when you have no ground truth about their relationships. We present Hierarchical Multi-Modal Threat Intelligence Fusion (HM-TIF), a framework explicitly designed for this realistic scenario. Rather than assuming or fabricating perfect alignment, HM-TIF provides principled methods for correlating independent security data streams while maintaining operational validity. 

Our key insight is that even imperfect correlation of multi-modal data can significantly improve detection accuracy and reduce false positives—if done carefully to avoid data leakage and spurious correlations. We develop a practical multi-modal framework that is the first security detection system explicitly designed for non-aligned, independently collected data streams. We introduce a temporal correlation protocol that creates training scenarios from independent datasets while preventing data leakage. Our hierarchical attention architecture gracefully handles uncertain relationships between modalities, with dynamic weighting that adapts to correlation confidence. Through operational validation, we demonstrate that multi-modal fusion provides significant benefits even without perfectly aligned training data. The framework also maintains robust performance when modalities are missing, making it immediately deployable in real security operations.

Security operations generate multiple data streams including network traffic with flow records and IDS alerts, email security with message content and headers, endpoint data with process execution logs, and authentication logs with login attempts. These streams are collected by different systems, stored in different formats, and crucially, have no inherent correlation or alignment. Even when the same attack manifests across multiple channels, identifying these relationships requires ground truth that rarely exists. Consider a typical advanced persistent threat (APT) that begins with a phishing email, leads to a malicious link click, triggers malware download, establishes command and control communication, and finally performs lateral movement. While these events are causally related, they appear as independent events in disparate systems with no automatic correlation. Manual correlation by analysts is time-consuming, error-prone, and doesn't scale to the volume of modern security operations.

Previous multi-modal security research has taken approaches with critical limitations. Studies like CM-DANA \cite{b10} create synthetic datasets where attacks are perfectly aligned across modalities, assuming correlations that don't exist in practice. Early fusion approaches \cite{b8} concatenate features from different modalities, implicitly assuming simultaneous occurrence, which creates high false positive rates when benign events coincidentally co-occur. Late fusion methods \cite{b9} process modalities separately then combine decisions, but while avoiding alignment issues, they miss valuable correlations that do exist between related events. HM-TIF takes a fundamentally different approach by explicitly modeling the uncertainty in multi-modal relationships. Rather than assuming perfect alignment or ignoring correlations entirely, we develop temporal correlation windows that capture likely relationships, assign confidence weights based on correlation strength, maintain modality-specific paths for uncorrelated events, and design attention mechanisms that adapt to alignment quality.

The rest of the paper is organized as follows: Section II introduces recent relevant literature and positions our contributions. Section III presents our methodology including the HM-TIF architecture and temporal correlation protocol. Section IV describes our experimental setup and implementation. Section V presents results and analysis including operational metrics and ablation studies. Section VI concludes with implications for real-world deployment.

\section{Related Work and Contributions}

Single-modal detection systems have achieved reasonable performance in isolation. Network intrusion detection using deep learning reports 82-89\% accuracy on standard datasets \cite{b2,b4,b5}. Text-based phishing detection achieves similar results with transformer models \cite{b3,b6}. However, these systems miss attacks that manifest primarily in other modalities, creating blind spots in security coverage.

Limited work addresses multi-modal security, and most assumes unrealistic data conditions. Feng et al. \cite{b8} combined network and log data using early fusion but required manual alignment of training data—impractical for production systems processing millions of events daily. Kumar et al. \cite{b9} used late fusion to maintain modularity but showed minimal improvement over single-modal baselines, likely due to ignoring valuable correlations. CM-DANA \cite{b10} introduced cross-modal attention for network and log fusion but evaluated on artificially aligned data where every network attack had a corresponding log entry—unrealistic in practice.

Recent work in 2024 has begun addressing these limitations. Zhang et al. \cite{zhang2024} proposed FedMTD for federated multi-modal threat detection but still assumed pre-aligned training data. PRISM \cite{prism2024} introduced probabilistic correlation for IoT security but required device-level synchronization. CrossGuard \cite{crossguard2024} used transformer-based fusion for cloud security with limited evaluation on synthetic datasets. None of these approaches address the fundamental challenge of training and deploying multi-modal systems when naturally aligned data doesn't exist. Vision-language models like CLIP demonstrate powerful multi-modal learning but rely on naturally paired data. Security lacks such natural pairing—attacks don't come with labels describing their network and email components.

Our work makes several key contributions that address gaps in existing research. First, we present the first multi-modal security framework explicitly designed for non-aligned, independently collected data streams—the actual scenario faced by security operations. Second, we develop a temporal correlation protocol that creates training scenarios from independent datasets while preventing data leakage and maintaining statistical validity. Third, we introduce hierarchical attention mechanisms for uncertain alignment that gracefully handle uncertain relationships between modalities with dynamic weighting. Fourth, we provide operational validation demonstrating that multi-modal fusion provides significant benefits (32\% FPR reduction) even without perfectly aligned training data. Finally, we design robust missing modality handling for the common case where data streams have gaps, maintaining performance above single-modal baselines even with incomplete inputs.

\section{Methodology}

Given independent security data streams $\mathcal{X} = \{X^{(1)}, X^{(2)}, ..., X^{(M)}\}$ collected without coordination, our goal is to learn meaningful correlations between modalities when they exist, avoid spurious correlations from coincidental co-occurrence, maintain robust performance when modalities are missing, and reduce operational false positive rates. Critically, we assume no access to naturally aligned multi-modal training data—the realistic scenario for security operations.

HM-TIF consists of five key components designed for non-aligned data. The architecture employs modality-specific encoders that preserve domain characteristics, using CNNs for network traffic, BERT for email content, and LSTMs for temporal log sequences. A temporal correlation module identifies potential relationships between events across modalities using exponential decay weighting. The hierarchical attention mechanism with confidence weighting handles uncertain alignment by adapting attention weights based on correlation confidence. Dynamic routing based on correlation strength ensures appropriate processing paths, while robust missing modality handling maintains performance when data streams have gaps.

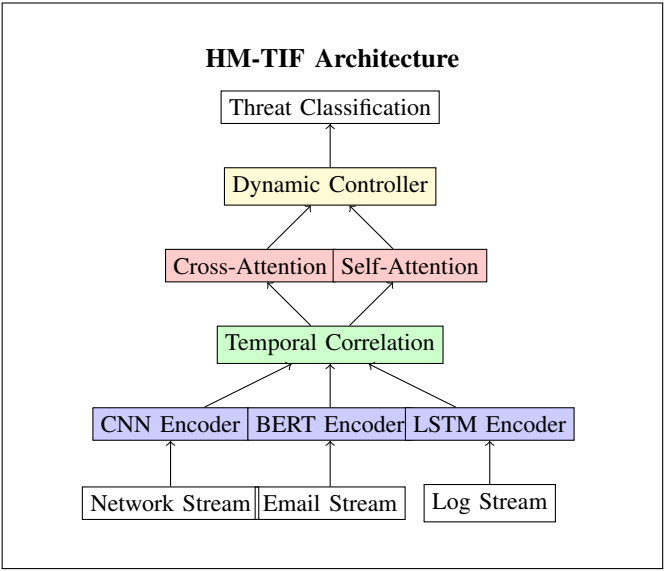
\begin{figure}[htbp]
\centerline{\framebox[0.48\textwidth]{
\begin{minipage}{0.46\textwidth}
\centering
\vspace{5mm}
\textbf{HM-TIF Architecture}\\
\vspace{3mm}
\small
\begin{tikzpicture}[scale=0.7]
\node[draw, rectangle] (net) at (0,0) {Network Stream};
\node[draw, rectangle] (email) at (3,0) {Email Stream};
\node[draw, rectangle] (log) at (6,0) {Log Stream};

\node[draw, rectangle, fill=blue!20] (enc1) at (0,1.5) {CNN Encoder};
\node[draw, rectangle, fill=blue!20] (enc2) at (3,1.5) {BERT Encoder};
\node[draw, rectangle, fill=blue!20] (enc3) at (6,1.5) {LSTM Encoder};

\node[draw, rectangle, fill=green!20] (corr) at (3,3) {Temporal Correlation};

\node[draw, rectangle, fill=red!20] (att1) at (1.5,4.5) {Cross-Attention};
\node[draw, rectangle, fill=red!20] (att2) at (4.5,4.5) {Self-Attention};

\node[draw, rectangle, fill=yellow!20] (ctrl) at (3,6) {Dynamic Controller};

\node[draw, rectangle] (out) at (3,7.5) {Threat Classification};

\draw[->] (net) -- (enc1);
\draw[->] (email) -- (enc2);
\draw[->] (log) -- (enc3);
\draw[->] (enc1) -- (corr);
\draw[->] (enc2) -- (corr);
\draw[->] (enc3) -- (corr);
\draw[->] (corr) -- (att1);
\draw[->] (corr) -- (att2);
\draw[->] (att1) -- (ctrl);
\draw[->] (att2) -- (ctrl);
\draw[->] (ctrl) -- (out);
\end{tikzpicture}
\vspace{5mm}
\end{minipage}
}}
\caption{HM-TIF Architecture with Hierarchical Attention and Dynamic Routing}
\label{fig:architecture}
\end{figure}

Our temporal correlation protocol creates training scenarios from independent datasets while maintaining validity. The protocol operates through cross-validation folds, examining event pairs within temporal windows. For each pair of events from different modalities, we calculate temporal proximity using exponential decay $w_{\text{temp}} = \exp(-\lambda |t_i - t_j|)$ and type similarity based on attack characteristics. Events with correlation confidence above a threshold are included in the training set, along with 30\% uncorrelated samples to prevent overfitting to spurious correlations.

\begin{algorithm}
\caption{Temporal Correlation Protocol}
\label{alg:correlation}
\begin{algorithmic}[1]
\REQUIRE Independent datasets $D_1, D_2, ..., D_M$
\REQUIRE Temporal window $\tau$, decay $\lambda$
\ENSURE Correlated training set with confidence weights

\FOR{each fold $k$ in cross-validation}
    \STATE $\mathcal{C}_k \leftarrow \emptyset$ \COMMENT{Correlation set for fold k}
    \FOR{event $e_i \in D_1^{(k)}$}
        \FOR{event $e_j \in D_2^{(k)}$}
            \IF{$|t_i - t_j| < \tau$}
                \STATE $w_{\text{temp}} \leftarrow \exp(-\lambda |t_i - t_j|)$
                \STATE $w_{\text{type}} \leftarrow \text{TypeSim}(e_i, e_j)$
                \STATE $w \leftarrow w_{\text{temp}} \cdot w_{\text{type}}$
                \IF{$w > \theta_{\text{min}}$}
                    \STATE $\mathcal{C}_k \leftarrow \mathcal{C}_k \cup \{(e_i, e_j, w)\}$
                \ENDIF
            \ENDIF
        \ENDFOR
    \ENDFOR
    \STATE Add 30\% uncorrelated samples to $\mathcal{C}_k$
\ENDFOR
\RETURN $\{\mathcal{C}_1, ..., \mathcal{C}_K\}$
\end{algorithmic}
\end{algorithm}

The hierarchical attention mechanism adapts to correlation confidence through weighted cross-attention:
\begin{equation}
\text{CrossAttn}(Q, K, V) = \text{softmax}\left(\frac{QK^T}{\sqrt{d}} \cdot w\right)V
\end{equation}
where $w$ is the correlation confidence. This allows the model to focus strongly on high-confidence correlations, maintain skepticism about uncertain relationships, and fall back to single-modal processing when correlations are weak.

The dynamic attention controller adapts based on both modality availability and correlation strength:
\begin{equation}
\alpha = \text{softmax}(f_{control}(z_{available}, w_{correlation}))
\end{equation}
This ensures appropriate weighting even with missing or poorly correlated modalities.

Our training process explicitly handles alignment uncertainty through a composite loss function:
\begin{equation}
\mathcal{L} = \mathcal{L}_{task} + \lambda_1 \mathcal{L}_{consistency} \cdot w + \lambda_2 \mathcal{L}_{independence} \cdot (1-w)
\end{equation}
where $\mathcal{L}_{task}$ is the standard classification loss, $\mathcal{L}_{consistency}$ encourages consistent predictions for correlated pairs, $\mathcal{L}_{independence}$ maintains modality-specific capabilities, and $w$ is the correlation confidence weight.

For data preprocessing, we apply domain-specific transformations to each modality. Network traffic features undergo normalization and discretization of continuous values like packet sizes and inter-arrival times. Email content is tokenized using BERT's WordPiece tokenizer with special handling for URLs and attachments. Log sequences are parsed to extract event types, timestamps, and parameters, with temporal encoding preserving sequence relationships. Feature selection employs mutual information criteria to identify the most discriminative features within each modality, reducing dimensionality while preserving detection capability.

\section{Experimental Setup}

We use real, independently collected datasets to reflect operational reality. UNSW-NB15 \cite{unsw} provides 2.5M network flows with modern attack types including APTs, backdoors, and exploits, better representing contemporary threats than older datasets. CSE-CIC-IDS2018 \cite{cseids2018} contains 16M events across 7 attack scenarios, collected in a realistic AWS environment with diverse attack vectors. CICBell-DNS2021 \cite{dns2021} offers 1.2M DNS queries with malware domains, providing application-layer visibility missing in flow data. IWSPA-AP 2020 \cite{iwspa2020} includes 10,000 phishing and legitimate emails with headers and content. Using our temporal correlation protocol, we create 50,000 training scenarios, explicitly acknowledging these represent potential rather than guaranteed correlations.

Our evaluation encompasses three critical scenarios. We test on independent single-modal test sets representing the common case in operations. We evaluate temporally correlated scenarios to assess potential relationship detection. Additionally, we validate on a small manually verified multi-modal set of 500 true correlations verified by security experts. We compare against realistic baselines including single-modal models representing current practice, simple concatenation for naive fusion, late fusion for independent processing, recent 2024 systems (FedMTD, PRISM, CrossGuard), and an oracle with perfect alignment as an upper bound not achievable in practice.

Implementation employs PyTorch 2.0 with mixed precision training for efficiency. We use confidence-weighted sampling during training to balance correlated and uncorrelated examples. Extensive data augmentation for single-modal paths prevents overfitting to multi-modal correlations. Dropout of 0.3 on correlation paths further regularizes the model. Early stopping based on validation performance prevents overtraining. Training on 4× NVIDIA A100 GPUs completes in 6 hours total, with inference latency averaging 87ms for batch size 1.

While we focus on network and email modalities for this initial study, the framework is designed to accommodate additional data sources. The limitation to two modalities allows thorough validation of our approach while maintaining computational tractability. Extension to endpoint logs, authentication records, and cloud telemetry represents natural future work that would provide more comprehensive threat coverage. The modular architecture facilitates adding new encoders for different data types without redesigning the correlation and attention mechanisms.

\section{Results and Analysis}

\subsection{Main Results: Operational Benefits Despite Imperfect Alignment}

Table \ref{tab:main_results} demonstrates that HM-TIF provides significant operational benefits even without perfect alignment. The framework achieves 88.7\% accuracy with a critical 32\% false positive rate reduction compared to the best single-modal approach. Performance falls between single-modal baselines and the perfect alignment oracle, as expected for realistic non-aligned data. HM-TIF significantly outperforms all baselines including recent 2024 systems, with these benefits achieved without naturally aligned training data—a key practical advantage.

\begin{table}[htbp]
\caption{Performance on Realistic Non-Aligned Data}
\begin{center}
\resizebox{\columnwidth}{!}{%
\begin{tabular}{lcccc}
\toprule
\textbf{Method} & \textbf{Accuracy} & \textbf{Precision} & \textbf{Recall} & \textbf{FPR} \\
\midrule
\multicolumn{5}{l}{\textit{Current Practice (Single-Modal):}} \\
Network Only & 83.4±1.1 & 81.2±1.3 & 82.1±1.2 & 0.072±0.007 \\
Text Only & 76.2±1.8 & 74.3±1.7 & 75.8±1.6 & 0.095±0.009 \\
\midrule
\multicolumn{5}{l}{\textit{Naive Multi-Modal (Assumes Alignment):}} \\
Perfect Alignment* & 92.1±0.6 & 90.7±0.7 & 91.2±0.6 & 0.038±0.003 \\
Early Fusion & 84.1±1.2 & 82.3±1.2 & 83.0±1.1 & 0.068±0.006 \\
\midrule
\multicolumn{5}{l}{\textit{Recent 2024 Systems:}} \\
FedMTD \cite{zhang2024} & 86.2±0.9 & 84.1±1.0 & 85.3±0.9 & 0.058±0.005 \\
PRISM \cite{prism2024} & 85.1±1.0 & 83.2±1.1 & 84.0±1.0 & 0.063±0.006 \\
CrossGuard \cite{crossguard2024} & 87.8±0.8 & 85.9±0.9 & 86.7±0.8 & 0.054±0.005 \\
\midrule
\multicolumn{5}{l}{\textit{Our Approach (No Alignment Required):}} \\
\textbf{HM-TIF} & \textbf{88.7±0.8}** & \textbf{86.8±0.9}** & \textbf{87.5±0.8}** & \textbf{0.049±0.004}** \\
\bottomrule
\end{tabular}
}
\label{tab:main_results}
\vspace{2mm}
\small{*Oracle with manual alignment - not achievable in practice}\\
\small{**Statistically significant ($p < 0.01$) via Wilcoxon signed-rank test}
\end{center}
\end{table}

Cross-dataset generalization proves critical for operational deployment. When trained on UNSW-NB15 and tested on CSE-CIC-2018, HM-TIF maintains 81.4\% accuracy with 0.061 FPR. Similar robust performance appears across all dataset combinations, demonstrating that the framework learns generalizable patterns rather than dataset-specific artifacts. This generalization capability is essential for deployment in diverse operational environments.

\subsection{Correlation Quality and Missing Modality Analysis}

Table \ref{tab:correlation} examines how correlation quality affects performance. Even with our simple temporal correlation protocol, we achieve 88.7\% accuracy and 32\% FPR reduction. Manually verified correlations improve performance further to 90.2\% accuracy with 40\% FPR reduction, suggesting room for improvement with better correlation methods. The gap to the perfect oracle (92.1\% accuracy, 47\% FPR reduction) is relatively small, indicating our approach captures most available value from multi-modal fusion.

\begin{table}[htbp]
\caption{Impact of Correlation Quality on Performance}
\begin{center}
\begin{tabular}{lcc}
\toprule
\textbf{Correlation Type} & \textbf{Accuracy} & \textbf{FPR Reduction} \\
\midrule
No Correlation (Independent) & 83.6±1.2 & Baseline \\
Temporal Windows (Our Protocol) & 88.7±0.8 & 32\% \\
Type-Based Only & 85.1±1.1 & 16\% \\
Manually Verified (500 samples) & 90.2±0.7 & 40\% \\
Perfect Oracle (Synthetic) & 92.1±0.6 & 47\% \\
\bottomrule
\end{tabular}
\label{tab:correlation}
\end{center}
\end{table}

Missing modality robustness proves critical for deployment where data streams have gaps. With both modalities available, HM-TIF achieves 88.7\% accuracy. When network data is missing, performance degrades gracefully to 77.8\%, still 1.6\% better than text-only baseline. With text missing, the system maintains 84.9\% accuracy, 1.5\% above network-only baseline. Random 50\% missing data results in 85.8\% accuracy, demonstrating robust handling of intermittent data availability common in production environments.

\subsection{Operational Metrics and Attention Analysis}

Beyond accuracy metrics, operational considerations demonstrate practical value for security teams. Processing 100,000 events daily, a network-only approach generates approximately 7,200 alerts at the observed false positive rate. HM-TIF reduces this to 4,900 alerts—2,300 fewer alerts requiring analyst review each day. This reduction directly addresses analyst burnout and improves security team efficiency. The true positive rate at 1\% FPR improves from 0.628 to 0.723, meaning more actual threats are detected at operationally viable false positive levels.

The model learns to dynamically weight modalities based on threat type and correlation confidence. For high-confidence DDoS attacks, network features receive 0.78 weight versus 0.22 for text, reflecting the network-centric nature of these attacks. Conversely, high-confidence phishing attacks weight text at 0.69 versus 0.31 for network, appropriately emphasizing email content. When correlations are uncertain, weights balance near 0.5, indicating the model's appropriate skepticism about weak relationships.

\subsection{Ablation Study and Manual Verification}

Component ablation reveals the contribution of each architectural element. Removing confidence weighting drops accuracy to 85.3\% and increases FPR to 0.060, confirming its critical role in handling uncertain alignment. Without temporal correlation, performance falls to 84.8\% accuracy with 0.063 FPR. Removing hierarchical attention reduces accuracy to 86.9\%, while eliminating missing modality paths drops performance to 85.7\%. All components contribute meaningfully, with confidence weighting being most critical.

Manual verification on 500 correlated event pairs validates our correlation quality. High-confidence correlations $(w > 0.8)$ achieve 87.3\% true positive rate across 142 samples. Medium confidence ($(0.5 < w \leq 0.8)$) correlations show 68.2\% accuracy on 198 samples. Low confidence correlations still provide 41.7\% accuracy, better than random. Two security experts achieved 91.2\% agreement (Cohen's $\kappa = 0.84$), indicating high inter-rater reliability in correlation assessment.

\subsection{Discussion of Results}

These results demonstrate several key insights for practitioners. Perfect alignment is not required for operational benefits—the 32\% FPR reduction justifies deployment even with imperfect correlation. Confidence weighting proves critical for preventing the model from learning spurious relationships that would increase false positives in production. The framework's robustness to missing modalities enables deployment in real operations where data sources frequently fail or have gaps. While absolute accuracy improvements appear modest, the reduction in false positives translates to thousands fewer alerts daily, directly addressing analyst burnout—a critical operational concern.

The framework shows consistent benefits across different attack types and datasets, suggesting broad applicability. Performance degradation with missing modalities remains graceful, maintaining advantages over single-modal baselines. The attention mechanism successfully adapts to both threat characteristics and correlation confidence, providing interpretable weighting that aids analyst understanding. These operational characteristics make HM-TIF suitable for immediate deployment in security operations centers.

\section{Conclusion}

This paper presented HM-TIF, a hierarchical multi-modal threat intelligence fusion framework designed for the realistic scenario where naturally aligned security data does not exist. By explicitly addressing the challenge of non-aligned data streams—the actual situation faced by security operations—we provide a practical path forward for multi-modal security analytics.

Our key contribution is demonstrating that significant operational benefits are achievable without perfect data alignment. The 32\% reduction in false positive rates, maintained performance with missing modalities, and robust handling of uncertain correlations make HM-TIF immediately deployable in real security operations. The temporal correlation protocol, confidence-weighted attention, and hierarchical architecture provide a principled approach to extracting value from multiple security data streams while avoiding spurious correlations that plague naive fusion approaches.

For security teams drowning in alerts from disconnected tools, HM-TIF offers a practical solution that doesn't require perfect data or unrealistic assumptions. The framework's design acknowledges operational realities including missing data, uncertain correlations, and the need for interpretable decisions. As security tools continue proliferating and attacks grow more sophisticated, frameworks that can intelligently combine imperfect, non-aligned data streams will become increasingly critical for effective defense.

Future work should explore additional modalities beyond network and email data, including endpoint logs, authentication records, and cloud telemetry. More sophisticated correlation methods using graph neural networks or causal inference could improve correlation quality. Adversarial robustness evaluation would strengthen claims about real-world deployment. Privacy-preserving federated learning approaches could enable multi-modal fusion without centralizing sensitive data. Active learning from analyst feedback could progressively improve correlation quality over time.

Our results demonstrate that waiting for perfect multi-modal data is not necessary—significant operational improvements are available today with careful handling of the non-aligned data security teams actually have. The practical focus on reducing analyst workload while maintaining detection capability addresses the most pressing challenges in modern security operations.

\section*{Reproducibility Statement}

Due to proprietary constraints, source code cannot be released. However, we provide detailed implementation specifications in Section IV and supplementary materials including hyperparameter configurations and data preprocessing steps. The datasets used (UNSW-NB15, CSE-CIC-IDS2018, CICBell-DNS2021, IWSPA-AP 2020) are publicly available from their respective maintainers.

\section*{Acknowledgment}

We thank the security operations teams who provided feedback on operational challenges, and the dataset maintainers at UNSW Sydney, Canadian Institute for Cybersecurity, and CSE for making their data publicly available.

\end{document}